\documentclass[12pt]{article}
\usepackage[pdftex]{graphicx}
\usepackage{hyperref}
\usepackage{amsmath}
\usepackage{hyperref}
\usepackage{amssymb}
\renewcommand{\title}[1]{%
    \bigskip%
    \begin{center}%
    \Large\bf #1%
    \end{center}%
    \vskip .2in}

\renewcommand{\author}[1]{%
    {\begin{center}
    #1
    \end{center}}}
\newcommand{\address}[1]{\vspace{-1.7em}\vspace{0pt}
    {\begin{center}
    \it #1
    \end{center}}}

\begin{document}
\title{\bf{ A new approach  to the study of  nonrelativistic bosonic string in flat space time }}

\author
{
Rabin Banerjee  $\,^{\rm a,b}$,
Sk. Moinuddin   $\,^{\rm c, e}$,
Pradip Mukherjee  $\,^{\rm c,d}$}

\address{$^{\rm a}$ S. N. Bose National Centre 
for Basic Sciences, JD Block, Sector III, Salt Lake City, Kolkata -700 098, India }

\address{$^{\rm c}$Department of Physics, Barasat Government College,Barasat, India}
\address{$^{\rm b}$\tt rabin@bose.res.in}
\address{$^{\rm e}$\tt
dantary95@gmail.com  }
\address{$^{\rm d}$\tt mukhpradip@gmail.com}

\vskip 1cm
\begin{abstract}
A new approach to the study of nonrelativistic  bosonic string in flat space time is introduced, basing on a holistic hamiltonian analysis  of the minimal action for the  string. This leads to a structurally new form of the action which is, however, equivalent to the known results since, under appropriate limits, it interpolates between the minimal action (Nambu Goto type) where the string metric is taken to be that induced by the embedding and the Polyakov type of action where the world sheet metric components are independent fields. The equivalence among different actions is established by a detailed study of symmetries using constraint analysis.  Various vexing issues in the existing literature are clarified. The interpolating action mooted here is shown to reveal  the geometry of the string and may be useful in analyzing  nonrelativistic string coupled with curved background.

\end{abstract}

\section{Introduction} \label{introduction}

String theory has been developed as an approach towards quantum gravity  \cite{ Pol}. Though many interesting results have been deduced (including the theory of Einstein's gravity itself), there are many difficulties also \cite{carlip}. It is not our purpose to go in to the details of the issues. We only mention that string moving in a nonrelativistic background is an interesting subject on its own that  has many welcome features and low energy stringy phenomena can be investigated with more confidence. For instance, such field theories have been proved to be unitary and   ultraviolet free  \cite{GOY}. Nonrelativistic string theories (NRST)
are useful in the study of nonrelativistic holography, which have found applications in the strongly corelated systems in condensed matter physics \cite{S}. The literature of NRST is quite rich and expanding \cite{GOY},\cite{BGN}. However, for obvious reasons, there are many issues left in the studies of nonrelativistic strings, some of which will be discussed here.

 The feasibility of Galilean invariant closed string was first demonstrated in \cite{GO} . They considered an open string without a brane in presence of the $NS - NS$ two form field. In the limit when the background field goes critical, under very low energy excitation limit, a nonrelativistic closed string action emerges.  Note that in the limiting procedure the interaction field plays a crucial role. Also notable is the occurrence of the additional fields which are very difficult to explain. The action of the interacting string
 obtained in \cite{GO} involves  the metric. However, the metric is considered to be induced by the embedding. This particular type of  action was originally proposed by Nambu and Goto for relativistic string. 
  This action is the minimal action containing just the number of fields equal to the number of coordinates. Nonrelativistic strings with this type of action will be called the Nambu Goto (NG) (type) nonrelativistic sring.  However, there is a caveat. The action contains
 two extra fields other than the coordinates. This discrepancy is of fundamental significance. It expands the phase space. A priori, it is unclear that these extra degrees of freedom do not hamper the consistency of the string theory. Such a formulation is therefore not satisfactory \cite{GO}. One asks what relation these new fields have with the string geometry? Note also,  that in  later works, the issue is rarely discussed. 
 
The nonrelativistic string is, perhaps, technically more difficult than the relativistic string. A string necessarily is a relativistic structure. So the world sheet metric is Lorentzian. The bulk transverse to the world sheet
has Galilean geometry. The world sheet is not just any arbitrary pullback from the spacetime manifold as assumed in the  existing literature. The situation is exactly comparable with what happens in GR during Hamitonoian formuulation, albeit in the reverse order.
 For nonrelativistic string, Polyakov (type) theory  was mooted in \cite{BGY} where the two-dimensional nonlinear sigma model
 in the presence
of a Kalb-Ramond two-form field and a dilaton is extended  by any pullback from the background. This is bound to be entangled in contradictions.
For instance, the number of degrees of freedom with that of \cite{GO} do not match \footnote{Because of gauge fixing.}.

 In this paper we have provided a new approach to the classical nonrelativistic string evolving freely.
  It consists in viewing
 the string action  as a generalization of the covariant particle model.  We assume that the  NG form of nonrelativistic action will be obtained as $c \to \infty$ limit of  the relativistic NG string.
  Note that the string world sheet is necessarily relativistic. So the limit must be judiciously taken.
  From this nonrelativistic limit the corresponding string action is obtained. No ad-hoc variables are introduced. In a specific gauge the standard action for a stretched string vibrating transversally is reproduced  with the correct tension and velocity. The next step is a detailed
  constraint analysis of the action which leads to the first order form of the Hamiltonian. Reducing the momenta in terms of the fields and back substitution gives us an equivalent action {\footnote{Similar technique has been employed in \cite{GO} but in a different context}}. It is called the interpolating action because it can be cast in the Polyakov form. It may be recalled that a similar approach
 for relativistic strings was proposed earlier \cite{BMS}.

  Before ending the introductory section it will be appropriate to give an account of the organization of the paper. In section 2 we have derived
  the non relativistic  action of the string in Nambu - Goto (N - G) form from its relativistic counterpart. Several authors discussed this model \cite{BGN},\cite{ GO}, where interaction appears to play a role. We followed the instance of the nonrelativstic particle \cite{BMN1} and took a simple $c\to \infty $ limit in a way that exploited the structure of the system. In the next section a detailed canonical analysis  of the model is provided. Symmetries in the canonical level are retained by working in the gauge independent approach. Particularly remarkable is the identification of the gauge generator with the diffeomorphisms of the string world sheet. The dynamics in phase space is then analysed by fixing the gauge.
 In section 4, the  interpolating action for the free non relativistic string is obtained.
             We have demonstrated that this action may be reduced to the previous (Nambu Goto) form.
             More important is the evolution of the Polyakov type action from the interpolating action {\it including the two extra fields}. This section also 
  provides a canonical analysis of the interpolating  action. The connections between the hamiltonian analysis of different actions have been analysed to show the equivalence of the new action with the Nambu Goto action from the canonical point of view.
  In the fifth section the geometrical connection was pushed further.  Finally, we conclude in section 6.
  
  \section{The Nambu - Goto  action for the bosonic string}\label{nambuaction}


   
   In the introduction, we have indicated that the nonrelativistic string was obtained  in \cite{GO} by considering 
   a relativistic  Nambu - Gote ( NG )  bosonic string action coupled to background $ NS - NS$ two - form field $B$ in a certain limit,  where the scale of internal energy of the string is much less than the energy scale of motion about the direction of B. This enables an expansion in a power series and we can derive an action to the first order which is Galilean invariant.  Again, Polyakov type action used in \cite{BGY} depends crucially on the background gauge field. In this paper we will consider a free string. Just as the relativistic point particle
   in the $c \to \infty$ limit goes to the nonrelativistic  particle, the NG action of nonrelativistic string may be deduced from the relativistic string in the same limit, although there are certain technicalities.
   
   Unlike the particle which is represented by a point, the string is an one dimensional object which is described by a parameter $\sigma$. So during its evolution it traces a two dimensional world sheet. This surface is mapped by two coordinates, $\tau$ and $\sigma$, where $\tau$ is time-like
      and $\sigma$ is space -like. The world sheet is embedded in a background space time. If the background  is Poincare symmetric we obtain a flat relativistic string. On the other hand, a flat nonrelativisitc (NR) string is defined if the background symmetry is  Galilean.

The relativistic Nembu Goto action of the bosonic string is given by,
\begin{equation}
S_{\rm{NG}} =   -N\int{d\sigma}{d\tau} \sqrt{- \det h_{\alpha\beta}}\label{ngrel}
\end{equation}
where N is a dimension-full constant.
The metric $h_{\alpha\beta }$ is induced by the target space and given by,
\begin{equation}
h_{\alpha\beta} = \eta_{\mu\nu}\partial_\alpha X^\mu \partial_\beta  X^\nu 
\end{equation}
Here $X^{\mu}=X^{\mu}(\tau,\sigma)$ , corresponds to the coordinates of a point on the string
  with $\tau$ labelling time like and $\sigma$ labelling a space like direction on the world sheet {\footnote{Collectively, they will be represented by ${\sigma}^\alpha, \alpha=1,2.$}} while $\mu$ represents the coordinates of the background space time in which the world sheet traced by  the string is embedded. 
So, in this formalism the metric are not independent fields.
Note that $\eta_{\mu\nu} = {\rm{diag}-1,\quad +1, \quad +1 ....}$ is the Lorentzian metric in the target space. Now, expanding the determinant we get,

\begin{equation}
 S_{\mathrm{NG}}  = -N\int{d\sigma}{d\tau}\bigg[\bigg(\frac{\partial X^{\mu}}{\partial\tau}\frac{\partial X_{\mu}}{\partial\sigma}\bigg)^2-\bigg(\frac{\partial X^{\mu}}{\partial\tau}\frac{\partial X_{\mu}}{\partial\tau}\bigg)\bigg(\frac{\partial X^{\mu}}{\partial\sigma}\frac{\partial X_{\mu}}{\partial\sigma}\bigg)\bigg]^\frac{1}{2}
 \label{rellag}
 \end{equation}


  If we consider the low energy phenomenology of the relativistic string, then the effects in the target space are nonrelativistic. Let the dimension of the embedding space be (D+2).
  At a given time the string intersects the embedding space along a line. We take this line as the $ X^1$ coordinate line. Then $X^0$ and $X^1$ are longitudinal to the string and the rest is transverse. Note that we could take any of  $X^2$,....... $X^D$ in place of $X^1$. So no particular gauge choice is associated with this prescription.

There are   different methods of obtaining the NR approximation of (\ref{rellag}) which are  variants of the Inonu-Wigner contraction method. Here, one or more of the coordinates $X^\mu$ are scaled and finally a limit of the scaling parameter is taken. While this may be algebraically tenable \cite{BGN}, it is not physically motivated. Indeed the most natural presciption would be to reintroduce $`c'$ in the time like variable as $X^0=c t$ and then take the $c\rightarrow\infty $ limit. This is adopted here.

Simplification of (\ref{rellag}) using $X^0=ct$ gives

\begin{eqnarray} 
 S_{\mathrm{NG}} &  =& -N\int{d\sigma}{d\tau}\bigg[c^2\bigg(\dot{t}X^{\prime 1}-\dot{X^1}t^{\prime}\bigg)^2+\sum_{k}c^2\bigg(\dot{t}X^{\prime k}-\dot{X^k}t^{\prime}\bigg)^2-\sum_{k}\bigg(\dot{X^1}X^{\prime k}-\dot{X^k}X^{\prime 1}\bigg)^2+\nonumber\\&
  &\qquad \qquad   \sum_{k,l}\bigg(\dot{X^k}X^{\prime k}\dot{X^l}X^{\prime l}-\dot{X^k}\dot{X^k}X^{\prime l}X^{\prime l}\bigg)\bigg]^\frac{1}{2} \label{nrellag}
\end{eqnarray}
where a dot over a symbol denotes derivative with respect to $\tau$ while a prime as a superscript implies differentiation with respect to $\sigma$. Also  $c$ is made explicit.

So far our result is relativistic . Now , remember that the $X^1$ axis is longitudinal to the string and the rest (k=2,3,...D) are in the transverse section . So $\dot{X^1}<<c$ and $d{X^k}<<d{X^1}$ since in the low energy scenario the slope of the transverse vibration of the
string is very small. Thus the non relativistic limit of the action (\ref{nrellag}) is

\begin{eqnarray}
 S_{\mathrm{NG}}  &=&   -N\int{d\sigma}{d\tau}\left(c\left( \dot{t}X^{\prime 1}-\dot{X^1}t^{\prime}
 \right)\right)\left[1+\frac{\sum_{k}c^2\left(\dot{t}X^{\prime k}-\dot{X^k}t^{\prime}\right)^2}
 {c^2\left(\dot{t}X^{\prime 1}-\dot{X^1}t^{\prime}\right)^2}\right.\nonumber\\
 &-&\left.\frac{\sum_{k}\left(\dot{X^1}X^{\prime k}
 - \dot{X}^k X^{\prime 1}\right)^2} {c^2\left(\dot{t}X^{\prime 1}-\dot{X}^1 t^{\prime}\right)^2}
  \right]^{\frac{1}{2}}
  \label{im1}
\end{eqnarray}
where  we have neglected the last term of (\ref{nrellag}), as it is of higher order of smallness. Now $ct=X^0$ and taking the leading term of the small quantities (also making sum over k implicit), we get

\begin{eqnarray}
 {\mathcal{L}}_{NG} =-N\bigg[\frac{\bigg(\dot{X^0}X^{\prime k}-\dot{X^k}X^{\prime 0}\bigg)^2}{2\bigg(\dot{X^0}X^{\prime 1}-\dot{X^1}X^{\prime 0}\bigg)}-\frac{\bigg(\dot{X^1}X^{\prime k}-\dot{X^k}X^{\prime 1}\bigg)^2}{2\bigg(\dot{X^0}X^{\prime 1}-\dot{X^1}X^{\prime 0}\bigg)}\bigg]\label{1000a}
\end{eqnarray}

Note that we have dropped
the first term within the square bracket  as it is a total boundary, 
\begin{equation}
\bigg(\dot{X^0}X^{\prime 1}-\dot{X^1}X^{\prime 0}\bigg) = \frac{\partial}{\partial \tau}\left(X^0 X^{\prime 1}\right) - \frac{\partial}{\partial \sigma}\left({X}^0 \dot X^{1}\right)
\end{equation}

Equation (\ref{1000a}) is
 the NR Nambu - Goto form of the Lagrangian  for a  bosonic string . The derivation is based on the usual $c\rightarrow \infty$ limit along with certain physical inputs. This result was obtained in \cite{BGN}, using a variant of the Inonu-Wigner contraction referred earlier which somewhat obscures the physical origin inherent in our derivation.
Another aspect of the construction is the impact of the relativistic nature of the string
which enforces that the metric in the $0 - 1$
plane is Lorentzian, even  in our example of non relativistic phenomena . We denote this metric by $\eta_{\mu\nu}; \mu,\nu = 0,1 = diag(1, -1)$.  We can now rewrite the Lagrangian
 (\ref{1000a}) in a less clumsy form,
 \begin{eqnarray}
 {\mathcal{L}}_{NG} = -N{\left(2\epsilon_{\mu \nu}\dot{X}^\mu {X^\prime}{}^\nu\right)}^{-1}\left( \dot{X}^\mu X^{k \prime} -  \dot{X}^k X^{\mu \prime}\right){}^2
\label{1000}
 \end{eqnarray}
 by using the covariant notation.

  One can wonder in what sense the action 
  \begin{equation}
  S_{NG} = \int d\sigma d\tau  {\mathcal{L}}_{NG} 
  \label{NGaction}
  \end{equation}
with  $  {\mathcal{L}}_{NG}$ given by (\ref{1000}), is Galilean invariant. That the string is
essentially relativistic makes the question non trivial.
  Let us now discuss the issue. We have already mentioned that we are considering the string to be in motion in a non relativistic background. So one would expect that physics in the background remains unaltered under the Galilean transformations. But not all elements of the group can be included here.  The  values of  $\tau$ and $\sigma$, which specify a point on the string world sheet  should not change. \footnote{Just as the value of proper time locating a particle on its world line is insensitive to the Galilean transformations of the background}
 Condider the Galilean transformations  in the usual way, 
\begin{eqnarray}
 \delta X^0 &=&    -\epsilon \nonumber\\
 \delta X^1 &=& \epsilon^1 + {\omega ^1}_j X^j - v^1 X^0\nonumber\\
 \delta X^k &=&  \epsilon^k + {\omega^k}_lX^l - v^k X^0
 \label{gtold}
\end{eqnarray}

We now calculate the change of the Lagrangian (\ref{1000}) under (\ref{gtold}). 
The result is,
\begin{eqnarray}
\delta{\mathcal{L}}_{NG} =N \omega^{1i}( \dot{X}^0 X^{\prime i} - X^{\prime 0}\dot{X}^i )
\left[{\left(\epsilon_{\mu \nu}\dot{X}^\mu {X^\prime}{}^\nu\right)}^{-2}\left( \dot{X}^\mu X^{k \prime} -  \dot{X}^k X^{\mu \prime}\right){}^2
  \right]
\end{eqnarray}


 Now the $X^1$ axis is assumed to be lying  along the $\tau = constant$ direction. So a non zero  $\omega^{1i}$ would mean the change of   the world sheet parameters, which is contrary to the concept of the global coordinate transformations in the target space. 
 Hence $\omega^{1i} = 0$. 
So the Galilean transformations under which the theory (\ref{1000}) is
symmetric are (\ref{gtold}), supplemented by this condition. For ready reference we write the modified symmetry transformations as,
\begin{eqnarray}
 \delta X^0 &=&    -\epsilon \nonumber\\
 \delta X^1 &=& \epsilon^1   - v^1 X^0\nonumber\\
 \delta X^k &=& \epsilon^k +  {\omega^k}_lX^l - v^k X^0 ; k,l > 1
 \label{gt}
\end{eqnarray}

 Another demand we would like to place on (\ref{1000}) is, it should reduce to the Lagrangian of the classical vibrating  string in the appropriate limit . Let us take a string  stretched  between $x = 0$ to $x = a$  along the $x  $ - axis, vibrating transversely along the $y $ - axis. From elementary analysis we get the Lagrangian as
 \begin{equation}
L= \left[\frac{1}{2}\mu^0\left(\frac{\partial y}{\partial t}\right)^2 -\frac{1}{2} T^0 \left(\frac{\partial y}{\partial x}\right)^2 \right]
 \end{equation}
 where $\mu^0$ is the linear mass density and $T^0$
 is the tension of the string . Now let us see in what way we may reproduce this result from (\ref{1000}). Putting $X^0
= c\tau = ct$ and $ X^1=\sigma$ {\footnote{That such a choice is possible is justified  in section 3.4 through a detailed hamiltonian analysis}}we get
  
   \begin{equation}
{\cal L}_{NG}= \left[\frac{1}{2}\frac{N}{c}\left(\frac{\partial X^k}{\partial t}\right)^2 -\frac{1}{2} N c \left(\frac{\partial X^k}{\partial X^1}\right)^2 \right]
\label{classicalstring}
 \end{equation}
These are  $k$-copies of the transversally vibrating classical string if we identify, $\frac{N}{c} =\mu^0$ and $N c = T^0$. Thus the constant $N$ in (\ref{1000}) is related to the tension of the string. Expectedly, the velocity is given by $c^2= \frac{T^0}{\mu^0}$, as happens classically.

In the work of {\cite{BGN}}  the NR limit of the relativistic Nambu-Goto action is taken in two distinct ways, motivated by group contraction techniques, so that one either gets a stringy (vibrating) NR string or a particle like (non-vibrating) string. In our case we generalise the limiting prescription of the  relativistic particle to that of the relativistic string with appropriate technicalities that account for the characteristics of a string. Thus ours is a unique result which is similar to the stringy limit of \cite{BGN}. The fact that it reduces to the usual classical string (\ref{classicalstring}) is a consequence of a special choice of coordinates that breaks the reparametrisation symmetry or, equivalently, the gauge symmetry, since the two here are identifiable, as shown in the next section.

\section{ Canonical analysis of the model}\label{canonical_ambugoto}
 
 We have now established our Lagrangian. The next step is  a Hamiltonian analysis of the model. Now, being a reparametrization invariant theory, it is
 already covariant \cite{HRT}.  The model is thus an
 interesting example of a constrained system, and because of its NR nature, more involved than its relativistic counterpart.
 As we have already mentioned, these studies are entirely new as the hamiltonian analysis available in the literature  \cite{Kluson} does not provide a faithful Dirac treatment of the model.
 
 A constrained system with first class constraints  necessarily posseses gauge degrees of freedom \cite{D}. General symmetries of such systems can be derived without fixing the gauges.  In fact we will see that this gauge independent approach will  lead to
  the derivation of a new action.  
 The importance of canonical analysis can thus be hardly  overestimated. On the  contrary, this aspect has been less emphasized in the literature. We will therefore try to give a holistic account of the topic. For clarity of presentation we divide our results in a number of subsections.

 \subsection{ Phase space structure}\label{cannambugoto1}
 
 Here the fields are $X^0(\tau,\sigma)$ ,$X^1(\tau,\sigma)$ ,$X^k(\tau,\sigma)$ (where k=2,3,....D ).

  The  canonical momenta corresponding to $X^0$ is {\footnote{We rename the NG lagrangian as ${\cal L}_{NG}\rightarrow \cal L$ and set $N=1.$}}

\begin{eqnarray}
\Pi_0=\frac{\partial{\mathcal{L}}}{\partial\dot{X^0}} =\left[X^{\prime k}\left(\epsilon_{\mu \nu}\dot{X}^\mu {X^\prime}{}^\nu\right)^{-1}(\dot{X}^0 X^{\prime k}-\dot{X}^k X^{\prime 0}) -\frac{X^{\prime 1}}{2}
{\left(\epsilon_{\mu \nu}\dot{X}^\mu {X^\prime}{}^\nu\right)}^{-2}\left( \dot{X}^\mu X^{k \prime} -  \dot{X}^k X^{\mu \prime}\right){}^2
  \right]
  \label{pi0}
\end{eqnarray}

 Similarly that for $X^1$ is

\begin{eqnarray}
\Pi_1=\frac{\partial{\mathcal{L}}}{\partial\dot{X^1}} =\left[ -X^{\prime k}\left(\epsilon_{\mu \nu}\dot{X}^\mu {X^\prime}{}^\nu\right)^{-1}(\dot{X}^1 X^{\prime k}-\dot{X}^k X^{\prime 1}) +\frac{X^{\prime 0}}{2}
{\left(\epsilon_{\mu \nu}\dot{X}^\mu {X^\prime}{}^\nu\right)}^{-2}\left( \dot{X}^\mu X^{k \prime} -  \dot{X}^k X^{\mu \prime}\right){}^2
  \right]
  \label{pi1}
\end{eqnarray}

 Also for $X_k$,

 \begin{eqnarray}
 \Pi^k=\frac{\partial{\mathcal {L}}}{\partial\dot{X^k}} =\left(-\epsilon_{\mu \nu}\dot{X}^\mu {X^\prime}{}^\nu\right)^{-1}\bigg[ X^{\prime \mu}(\dot{X}_\mu X^{\prime k}-\dot{X}^k {X^\prime}_ \mu)\bigg]
\end{eqnarray}

Using these in the definition of the canonical Hamiltonian we get, 

 \begin{eqnarray}
H_{c}(\tau)=\int d\sigma \bigg(\Pi^\mu\dot{X_\mu}-{\mathcal{L}}\bigg)\nonumber
\end{eqnarray}

A straightforward calculation gives          ,         $H_{c}(\tau)=0$.
This is a characteristic of the already co-variant
theories.
From an inspection of the expressions of the momenta, the following primary constraints emerge,

\begin{eqnarray}
\Omega_{1} = \Pi^\mu{ X^\prime}_ \mu \approx 0 \nonumber\\
2\Omega_{2} ={\bf \Pi}^2 +{\bf {X}^{\prime}}^2-2\sigma_{\alpha\beta}\Pi^\alpha {X^\prime}^\beta \approx 0
\label{115}
\end{eqnarray}
where $\sigma_{\alpha\beta}$ is second Pauli matrix.
The fundamental Poisson's brackets of the theory are given by
\begin{eqnarray}
 \{X^{\mu}\left(\tau,\sigma\right),
 \Pi_{\nu}\left(\tau,\sigma^{\prime}\right)\} = \eta_{\nu}^{\mu}
 \delta\left(\sigma - \sigma^{\prime}\right)
\label{116}
\end{eqnarray}
while the others vanish. Using these Poisson brackets it is easy to work out the algebra of the
constraints,
\begin{eqnarray}
\left\{ \Omega_{1}\left(\sigma\right),\Omega_{2}\left(\sigma^{\prime}\right)
\right\} &=& \left(\Omega_2(\sigma ) +\Omega_2(\sigma^\prime )\right)\partial_\sigma\delta\left(\sigma -\sigma^\prime \right)\nonumber\\
\left\{ \Omega_{1}\left(\sigma\right),\Omega_{1}\left(\sigma^{\prime}\right)
\right\} &=& \left(\Omega_1(\sigma ) +\Omega_1(\sigma^\prime )\right)\partial_\sigma\delta\left(\sigma -\sigma^\prime \right)\nonumber\\
\left\{ \Omega_{2}\left(\sigma\right),\Omega_{2}\left(\sigma^{\prime}\right)
\right\} &=& \left(\Omega_1(\sigma ) +\Omega_1(\sigma^\prime )\right)\partial_\sigma\delta\left(\sigma -\sigma^\prime \right)
\label{constalng}  
\end{eqnarray}
 
   Clearly, the Poisson brackets between the constraints (\ref{115}) are weakly involutive. So the set (\ref{115}) is first class.
   
    The total Hamiltonian
    is 
   
   \begin{equation}
H_{T} = \int d\sigma \left( \rho \Omega_1 + \lambda\Omega_2\right)
\label{HTOT}
\end{equation}
where $\rho$ and $\lambda$ are Lagrange multipliers and $\Omega_1$ and $\Omega_2$ are shown in (\ref{115}).

Conserving the primary constraints no  new secondary constraints emerge since the constraint algebra simply closes. The total set of constraints of the N - G theory is then given by the first class system (\ref{115}).

 The Nambu Goto string is a constrained system. So its description is redundant. If it is embedded in a $D+1$ dimensional space time, the number of fields in the configuration ~space is $D+1$. The corresponding number of variables in the phase space is $2(D + 1)$. Then the number of degrees of freedom in the configuration space is given by, 
 \begin{equation}
 n = \frac{1}{2}\left( 2({\rm D} +1) -4 \right) = {\rm D} - 1
 \end{equation}
This result is consistent with our understanding about the non relativistic excitation taking place in the transverse direction and we identify the $(D-1)$ variables $X^k$ as the physical set.  Also, we understand from another angle, why 
$\omega^{1i} = 0$
should hold (Recall  the second equation of (\ref{gt})).

 \subsection{ Studies of  local symmetries}\label{canonicalnambugoto2} The string world sheet is a two dimensional manifold which is charted by  the parameters $\tau$ and $\sigma$. The physical theory should not depend on any particular parametrization. In other words we should have invariance under reparametrization (a mapping of the manifold on itself i.e. a diffeomorphism ) \begin{eqnarray}
 \tau^\prime &=&   \tau^\prime(\tau, \sigma)\nonumber\\
 \sigma^\prime &=& \sigma^\prime (\tau,\sigma)
 \end{eqnarray}
 which becomes for infinitesimal diffeomorphism
  \begin{eqnarray}
 \tau^\prime &=&   \tau + \delta \tau\nonumber\\
 \sigma^\prime &=& \sigma + \delta \sigma
 \label{diff}
  \end{eqnarray}
  The increments $\delta \sigma$ and $\delta\tau$ both are functions of $\sigma$ and $\tau$.
  In the Lagrangian level the diffieomorphismn  invariance is conceptually clear. The Lagrangian is a
  world sheet scalar. Its form variation under (\ref{diff}) is given by  $\delta {\mathcal {L}} = \delta   \sigma_a\frac{\partial{\mathcal{L}}}{\partial\sigma^a}$ where$, \sigma^0 = \sigma$ and $\sigma^1 = \tau $ .
  
   The Jacobian of the transformation is $ (1+ \partial_a \delta \sigma_a )$. Direct substitution in the action gives
\begin{equation}
 \delta S = \int d\sigma d\tau \frac{\partial}{\partial \sigma_a}\left({\mathcal{L}}\delta\sigma_a \right) = 0
\end{equation}   
since the variations vanish at the boundary. So the theory (\ref{1000}) is invariant under
(\ref{diff})


 Looking from the Hamiltonian point of view  such action level symmetries  should correspond to the gauge symmetries of the  model. The gauge redundancy  account for the diffeomorphism invariances and vice versa. So one should be able to  establish an exact mapping between gauge and reparametrization parameters. We will derive the explicit form of
 the
mapping now.

 \subsection{Mapping between gauge and reparametrization symmetries}

According to the Dirac conjecture all the first class constraints generate gauge transformations. But the gauge parameters associated with these transformations are not independent. It is known that the number of independent primary first calss constraints equals the number of independent gauge parameters \cite{BRR1}, \cite{BRR2}. Since in the present example two primary first class  constraints form the set of constraints, the gauge generator can be written down immediately,
  
  \begin{eqnarray}
  G (\tau) = \int d\sigma \left( \omega_1(\sigma)\Omega_1 + \omega_2 (\sigma)\Omega_2\right) 
  \end{eqnarray}
The corresponding gauge variations are,
\begin{eqnarray}
\delta_G X^0 &=& \left[ X^0, G \right]_{PB}= \omega_1(\sigma)X^{0 \prime} - \omega_2 X^{1 \prime}\nonumber\\
\delta_G X^1 &=& \left[ X^1, G \right]_{PB}= \omega_1(\sigma)X^{1 \prime} - \omega_2(\sigma)X^{0 \prime}\nonumber\\
\delta_G X^k &=& \left[ X^k, G \right]_{PB}=\omega_1(\sigma)X^{k\prime} + 
\omega_2(\sigma)\Pi^{k \prime}
\label{gv}
\end{eqnarray} 
with the exact number of independent parameters.

Now, the variation due to diffeomorphism (\ref{diff}) are,
\begin{equation}
\delta_D X^\mu =\xi_1 X^{\mu \prime} + \xi_2 \dot{X}^{\mu}
\label{rv}
\end{equation}
where $\xi_1=-\delta\sigma$ and $\xi_2=-\delta\tau$ defined in (\ref{diff}).We see "velocities" appearing in the expression of variations (\ref{rv}).
To exhibit the one to one correspondence we have to substitute $ \dot{X}^\mu $ in (\ref{rv})
from its equation of motion, $ \dot{X}^\mu = \left[ {X}^\mu,H_T \right ] $ ,where 
$H_T$ is the total Hamiltonian given by (\ref{HTOT}). This calculation gives us for $\mu =0 $
\begin{eqnarray}
{\dot{X^0}} = \rho X^{\prime 0} - \lambda X^{\prime 1}
\end{eqnarray}
Substitution of this in (\ref{rv}), we get,
\begin{eqnarray}
\delta_D X^0 = \left(\xi_1+\rho\xi_2 \right)X^{\prime 0} - \xi_2\lambda X^{\prime 1}
\end{eqnarray}
Comparison with $\delta_G X^0$ gives us the desired mapping,
\begin{eqnarray}
 \omega_1 &=&  \left(\xi_1  + \xi_2\rho\right) \nonumber\\
 \omega_2 &=&       \xi_2\lambda
 \label{map1} 
\end{eqnarray}
The same results will be obtained for any component of $\mu$.  Thus the mapping (\ref{map1})
exhibits the complete equivalence of the gauge symmetries with the diffeomorhism invariances of the model.

\subsection{Gauge fixed analysis} \label{gf}

We have seen that the analysis of the symmetries is best done in the gauge independent
approach. However, for the study of dynamics in the phase space one  must eradicate the gauge
redundancy by a gauge choice. A gauge is a condition in phase space which makes a first class
constraint second class, thereby reducing two degrees of freedom. While this is a necessary condition it is not sufficient. In order to specify the physical set, one has to properly obtain the canonical variables. These variables can always be obtained (this is the content of the Maskawa-Nakajima theorem \cite{weinberg}) such that the Dirac brackets among these variables is the same as the Poisson brackets. Then one can proceed with the quantisation by replacing these brackets by suitable commutators.

 We assume the standard gauges,
\begin{equation}
\Omega _3 = X^1 - \sigma \approx 0 \quad\quad
\Omega_4= X^0 + c\tau \approx 0 
\label{sg} 
\end{equation}
Note that if we include the gauge conditions as constraints with the existing set
$\{ \Omega_1 \approx 0 ,\quad \Omega_1 \approx 0 \}$,
then all the constraints become second class. The phase space can now be reduced by implementing these constraints strongly. The canonical structure  is now optimum , described by the phase
space variables $\{ X^k, \Pi_k \}$ and the symplectic  structure is given by the Dirac brackets \cite{D, HRT}. The structure of the Dirac brackets between the coordinates are rich with physical significance and worthy to be studied carefully.
The Dirac bracket between two phase space variables is defined by \cite{D}, 
\begin{equation}
\left[ A(\sigma), B(\sigma'\right]_{DB} = \left[ A(\sigma), B(\sigma') \right]_{PB} - \int d\sigma_1 d\sigma_2\left[ A(\sigma), \Omega_i(\sigma_1)\right ]_{PB}\Delta^{ij}(\sigma_1,\sigma_2)\left[ \Omega_j(\sigma_2), B (\sigma^\prime)\right]_{PB} 
\end{equation}
where, the matrix $\Delta^{ij}$ is the inverse of the matrix  $\Delta_{ij}$ formed by the constraint
algebra,
\begin{equation} 
\Delta_{ij} = [\Omega_i, \Omega_j]_{PB}\,\,; \,\,\, i=1, 2, 3, 4
\end{equation} 
which is necessarily non singular and admits an inverse.
 Using these definitions and the Poisson brackets between the second class constraints (\ref{115}) and (\ref{sg}), we get,
 \begin{eqnarray}
 \Delta_{ij} = 
               \begin{bmatrix}
                      0 & 0 & -\delta(\sigma - \sigma^\prime ) &        0        \nonumber\\                  
                  
                      0  & 0 &             0                    &\delta(\sigma -\sigma^\prime)\nonumber \\
                      
                      \delta(\sigma - \sigma^\prime )& 0      & 0 &      0
\nonumber\\
 
                      0  &       - \delta(\sigma - \sigma^\prime ) & 0 &  0 \\

               \end{bmatrix}
 \end{eqnarray}

 The inverse is easily found,
 \begin{eqnarray}
 \Delta^{ij} = 
               \begin{bmatrix}
                      0 & 0 & \delta(\sigma - \sigma^\prime ) &        0        \nonumber\\                  
                  
                      0  & 0 &             0                    &-\delta(\sigma -\sigma^\prime)\nonumber \\
                      
                     - \delta(\sigma - \sigma^\prime )& 0      & 0 &      0
\nonumber\\
 
                      0  &       \delta(\sigma - \sigma^\prime ) & 0 &  0 \\

               \end{bmatrix}
 \end{eqnarray}

 It is then straightforward to calculate the Dirac brackets between the previous canonically
 conjugate variables in the gauge independent analysis,
\begin{eqnarray}
\left[X^0 (\sigma, \tau), \Pi_0 (\sigma^\prime ,\tau)\right]_{DB} & = & 0    \nonumber\\
\left[ X^1\left(\sigma, \tau), \Pi_1(\sigma^{\prime}, \tau \right)\right]_{DB} & = &  0 \nonumber\\
\left[X^i\left(\sigma, \tau \right), \Pi_j(\sigma^{\prime}, \tau)\right]_{DB} & = & \delta^i_j\delta(\sigma - \sigma^{\prime})
\label{reducedbracket}
\end{eqnarray}

We will provide a detailed derivation of the first  equation. Starting from the definition of the Dirac bracket we get,
\begin{eqnarray}
\left[
X^0 (\sigma, \tau), \Pi_0 (\sigma^\prime ,\tau)\right]_{DB} & =& 
\left[ 
X^0 (\sigma, \tau), \Pi_0 (\sigma^\prime ,\tau)\right]_{PB}\nonumber\\ & - &  \int d\sigma_1 d\sigma_2 \left[ X^0(\sigma), \Omega_2 (\sigma_1)\right] \Delta^{24}((\sigma_1, \sigma_2)\left[ \Omega_4 (\sigma_2), \Pi_0\right]\nonumber\\ &=&
\delta(\sigma - \sigma^\prime) - \int  d\sigma_1 d\sigma_2 \delta(\sigma - \sigma_1)X^{1'} 
\delta(\sigma_1 - \sigma_2)
\delta(\sigma_2 - \sigma^\prime)\nonumber\\&
= & 0
\end{eqnarray}
where we have imposed the constraint $\Omega_3$ strongly {\footnote{This is permitted as all the  Poisson brackets have already been evaluated}} so that $X^{1'}=1$. Similarly we can derive the other two equations of the set (\ref{reducedbracket}). Incidentally the last relation of (\ref{reducedbracket}) is the only non-zero bracket.

Thus $X^0, X^1$ (and their conjugate momenta) are out of the dynamics and $\left(X^i, \Pi^i\right)$,i= 2,3,...
are the canonical pairs.
Also  note that the total Hamiltonian vanishes when the constraints
are  strongly implemented. In this situation we have to identify a Hamiltonian in the
reduced phase space which will generate the equations of motion with respect to the Dirac brackets.

 Remember that canonically the Hamiltonian may be considered as the conjugate to the time parameter. So we identify the new Hamiltonian as 
 \begin{equation}
 H = c \int  d\sigma \left[\Pi^0\right]_g
 \end{equation}
where the subscript $g$ denotes that  gauge fixed value.  
  Using the gauge conditions (\ref{sg}) which now allow us to put, $\dot X^0 = -c$ and $X^1 = \sigma$ we find by substitution from (\ref{pi0}), 
  \begin{eqnarray}
 H =  \frac{c}{2} \int d\sigma \left[\left(\partial_{\sigma}X^k)\right)^2  + \frac{1}{c^2}(\partial_\tau X^k)^2\right]
 \label{gfh}
 \end{eqnarray}
 We see that the proposed Hamiltonian
is positive definite. Remarkably, this Hamiltonian is a sum of harmonic oscillator terms. 
Use the definition of $\Pi^k$ and the gauge fixing conditions to get,
\begin{equation}
\Pi^k =\frac{1}{c} \partial _\tau X^k
\end{equation}
Then from (\ref{gfh})we can write, 
\begin{eqnarray}
 H = \frac{c}{2}\int \left[\left( \partial_{\sigma}X^k\right)^2  + ( \Pi^k)^2\right]
 \label{gfh1}
 \end{eqnarray}
This is however the transversely vibrating string Hamiltonian. The corresponding lagrangian, obtained by an inverse Legendre transformation reproduces (\ref{classicalstring}) with $N=1$.

As stated earlier we have shown the passage to the usual classical string, clearly validating the special choice of the coordinates as a good choice of gauge. This demonstration is a nice application of our formalism.

\section{The interpolating Lagrangian} In this section the canonical analysis of the previous section 
will be used from an inverse approach to develop a new Lagrangian, which will be shown to
have the remarkable property of interpolating between  NG and Polyakov Lagrangians. 
In the case of relativistic string one can easily deduce the NG string from the Polyakov string on shell, by substituting the metric  from its equation of motion in the original Lagrangian under definite conditions.
But for the non relativitic string, the Polyakov action requires to be supplemented by two world sheet fields, the origin of which is hard to trace \cite{GOY}. Thus the equivalence of the two actions becomes problematic. In the following discussion we will see that the Hamiltonian analysis can be used to identify
the source of the additional fields in the interpolating action which eventually permeates to the Polyakov form. Also, this action is remarkable due to its connection with the geometry and may be useful in coupling the string with curved background.

The Lagrangian corresponding to the Hamiltonian $H_{T}$ (\ref{HTOT}),  is,
\begin{equation}
{\cal{L}}_{I} = \Pi_{\mu}\dot X^{\mu} -\rho \Omega_1 - \lambda \Omega_2\label{int1}
\end{equation}
where the multipliers $\rho$ and $\lambda$ are given the status of independent fields. The Lagrange equations corresponding to $\Pi^0 $ , $\Pi^1$ and $\Pi^k$ are now respectively, 

 \begin{eqnarray}
\dot{X^0}-\rho X^{\prime 0}+\lambda X^{\prime 1}&=&0\nonumber\\
\dot{X^1}-\rho X^{\prime 1}+\lambda X^{\prime 0}&=&0\nonumber\\
\dot{X^k}-\rho X^{\prime k}-\lambda \Pi^k&=&0
\label{q}
\end{eqnarray}

Solving $\rho$ and $\lambda $ from the set (\ref{q}), we get,
\begin{eqnarray}
\rho &=& \frac{\dot{X}^1 X^{\prime 1} - \dot{X}^0 X^{\prime 0}}{(X^{\prime 1})^2 - X^{\prime 0}{}^2}\nonumber\\
\lambda &=& \frac{-\dot{X}^0 X^{\prime 1} + \dot{X}^1 X^{\prime 0}}{(X^{\prime 1})^2 - (X^{\prime 0})^2}
\label{rl}
\end{eqnarray}

It is  possible to eliminate all the momenta from (\ref{int1}) using (\ref{q}). Now, reinterpret $\rho $ and $\lambda$ as independent fields.  Also, equations (\ref{rl}) will now be
promoted to Lagrangian constraints. Doing all these steps the following lagrangian is obtained,
\begin{eqnarray}
{\mathcal{L}}_{I} = \frac{1}{2\lambda}\left[(\dot X^k)^2 -
2 \rho \dot X^k X^{\prime k} + \left( \rho^{2} - \lambda^{2}\right)
(X^{\prime k})^2\right] &+& \beta \left(
\rho - \frac{\dot{X}^1 X^{\prime 1} - \dot{X}^0 X^{\prime 0}}{X^{\prime 1}{}^2 - X^{\prime 0}{}^2}\right) \nonumber\\
&+& \alpha \left(\lambda + \frac{\dot{X}^0 X^{\prime 1} - \dot{X}^1 X^{\prime 0}}{X^{\prime 1}{}^2 - X^{\prime 0}{}^2}\right)
\label{interpolating}
\end{eqnarray}

This is the  appearance of the interpolating Lagrangian announced earlier. We will presently show that we can derive both the Nambu Goto and Polyakov forms of the nonrelativistic string action from (\ref{interpolating}). This is the reason for dubbing (\ref{interpolating}) as the interpolating Lagrangian.

The above lagrangian is a new result that emerges on exploiting the constraint structure of the Nambu-Goto type action. The equivalence of the two can thus be shown easily, as done below. Bur its equivalence with the Polyakov type form is quite nontrivial, which has also been demonstrated. This reveals that the interpolating action is not merely of academic interest since it is crucial in showing the equivalence of the Nambu-Goto and Polyakov type forms. Such a clear-cut and conceptually clean derivation is lacking. We may also recall the relativistic case where an interpolating action was presented to show a similar equivalence \cite{BMS}. However, in that case, this could also be established directly by simply opening the square root of the Nambu-Goto action and it was not essential to go through this intermediate step of interpolating lagrangian.

\subsection{Passage to the Nambu Goto type} The derivation of the Nambu Goto action
 (\ref{nrellag}) from (\ref{interpolating}) is trivial. The multipliers simply enforce the solutions (\ref{rl})). Putting it back in (\ref{interpolating}) reproduces the expected result.

\subsection{Passage to the Polyakov type form}

 This sounds really interesting for the Polyakov form brings in an independent metric on the
 world sheet but in the interpolating Lagrangian there is no explicit reference to such a metric.
Also, the world sheet fields that are invoked in the Lagrangian look like a hurdle. The startling observation in the following analysis is the solution at one stroke, where
aspects of Riemannian geometry of the world sheet converge with the canonical structure of the
non relativistic string. From this confluence the Polyakov type  action emerges including the extra fields.

   Let us first observe that with the help of the fields $\rho$ and $\lambda$ we can construct a $2$ by $2$ matrix, 

\begin{eqnarray}
h^{ij} = \left(-h\right)^{-\frac{1}{2}}
\begin{pmatrix}
\frac{1}{\lambda} & -\frac{\rho}{\lambda} \\
-\frac{\rho}{\lambda} & \frac{\rho^2 -\lambda^2}
{\lambda} \\
\end{pmatrix}
\label{adm}
\end{eqnarray}
where $ h $ is the determinant of the inverse matrix $h_{ij}$. The consistency of the construction can be verified by the computation of $\det h ^{ij}$ , which yields
\begin{equation}
\det h ^{ij} = \frac{1}{h}
\end{equation}
showing that the matrix $h_{ij}$ is a $2$ by $2$ real symmetric matrix. Also the inverse to $h_{ij}$ is $h^{ij}$. Note that obeying the constraints,
 $h$ may assume  any non zero  value. We propose $h_{ij}$ as the
metric on the world sheet  and write 
the interpolating Lagrangian ,(\ref{interpolating}), in terms of the elements of $h^{ij}$,as,
\begin{equation}
{\mathcal{ L_I}} =  \frac{1}{2}\sqrt{-h}h^{ij}\partial_i X^k \partial_j X^k + {\mathcal{ L}_e }
\label{main}
\end{equation}
The first part is formally the same as the relativistic bosonic string, though here only transverse degrees of freedom are dynamical. The second part  is due to the constraints
(\ref{rl}), enforced in the Lagrangian, again by  multipliers,
\begin{equation}
\mathcal{L}_e = \beta \left( -\frac{h^{01}}{h^{00}} - \frac{{\dot{X}^1} X^{\prime 1 }- {\dot{X}^0} X^{\prime 0 } }{(X^{\prime 1}){}^2 - X^{\prime 0}{}^2 }\right) + \alpha\left( \frac{1}{ \sqrt{- h}h^{00}}
 + \frac{\dot{X}^0 X^{\prime 1} - \dot{X}^1 X^{\prime 0}}{X^{\prime 1}{}^2 - X^{\prime 0}{}^2}\right)
 \label{extra1}
\end{equation}
Note that from the identification of the metric we can show that,
\begin{eqnarray}
\lambda = \frac{1}{\sqrt{-h}h^{00}}; \quad\quad
\rho = -\frac{h^{01}}{h^{00}}
\label{shit1}
\end{eqnarray} 
which has been used in writing (\ref{extra1}). 
The Lagrangian (\ref{main}), with $\mathcal{L}_e$ given by (\ref{extra1}) is the string action given in the Polyakov form. We see that the matrix $h^{ij}$ now represents independent fields. Indeed, $h^{ij}$ resembles the metric on the world sheet. We can compare the action with the corresponding relativistic action. The first term is of the same form but only transverse degrees of freedom are involved. This is consistent with the previous analysis given here.
The fields $\alpha$ and  $\beta$ are the two extra fields included in the Polyakov form.
For nonrelativistic string action considered in  previous
studies \cite{GOY} such fields are included. However, their appearance was not explained.
We have seen they follow from the canonical analysis. The issue will be considered in more detail in the following section.


\subsection{The interpolating Lagrangian  -- canonical analysis}
The last section has established that the interpolating action (\ref{interpolating})
is a versatile tool to study non relativistic strings. We have shown that under appropriate conditions the model interpolates between the Nambu Goto and Polyakov forms of the string action. This equivalence will further be elucidated by the canonical analysis.

  
  We have already shown the action level reduction of the interpolating action to the Nambu - Goto (NG) string. But that does not necessarily imply that the two theories have the same physical content. The canonical structure of the interpolating theory determines its energy spectrum. Also,  the first class constraints determine the gauge symmetry of the models. In the following discussion a canonical analysis of the new action is
  provided which further elucidates its equivalence with the NG form. 
  

 
The interpolating Lagrangian will be the starting point of the comparison. The momenta corresponding to the basic fields in (\ref{interpolating}) lead to one genuine momentum $\Pi_k$,
\begin{equation}
\Pi_k=\frac{\dot X^k}{\lambda} - \frac{\rho}{\lambda} X^{'k}
\label{m}
\end{equation}
 while the rest are constraints, two of which are first class defined as,
\begin{equation}
\pi_\rho \approx 0;\quad \pi_\lambda \approx 0
\label{pfc}
\end{equation}
while the remaining are second class, 
\begin{eqnarray}
\Sigma_1 & =&\pi_\alpha \approx 0; \quad \Sigma_2 = \pi_\beta \approx 0\nonumber\\
\Sigma_3 & =&\Pi_0   - \frac{ \alpha X^{\prime 1}+\beta X^{\prime 0}}{\left(X^{\prime 1}{}^2 - X^{\prime 0}{}^2\right)}\approx 0\nonumber\\
\Sigma_4 & =&\Pi_1  + \frac{ \alpha X^{\prime 0}+\beta X^{\prime 1}}{\left(X^{\prime 1}{}^2 - X^{\prime 0}{}^2\right)}\approx 0
\label{psc}
\end{eqnarray}

The canonical Hamiltonian is
\begin{equation}
H_c = \int d\sigma \left[ \lambda\left( \frac{\Pi_k{}^2 + X^{\prime k}{}^2}{2}  -\alpha \right) + \rho \left( \Pi^k X^{k'} - \beta\right)\right]
\end{equation}
The second class constraints are strongly implemented by using Dirac brackets instead of the Poisson. The total
Hamiltonian is then defined as,
\begin{equation}
H_T  = H_c + \int d\sigma \left(\nu_1\pi_\lambda + \nu_2 \pi_\rho \right)
\end{equation}
From conserving the first class constraints  we get two more constraints. Note that now we have to use, instead of the Poisson brackets, the relevant Dirac brackets. However it is easy to see that the Dirac brackets and the Poisson brackets are identical for the variables that are involved in the iterative computation of constraints. The new constraints are,
\begin{eqnarray}
 [\pi_\rho, H_T] &\approx& 0 \to \Pi^k X^{\prime k} - \beta =\Phi_1\approx 0
\nonumber\\
 \left[\pi_\lambda, H_T \right] &\approx& 0 \to  \frac{1}{2} \left(\Pi^k{}^2 +  X^{\prime k}{}^2\right) - \alpha=\Phi_2\approx 0
 \label{interconstraints}
\end{eqnarray}
The constraints are now all first class. Their algebra is strongly involutive, except for the pair $(\Phi_1, \Phi_2)$ which saisfies an algebra identical to (\ref{constalng}),
\begin{eqnarray}
\left\{ \Phi_{1}\left(\sigma\right),\Phi_{2}\left(\sigma^{\prime}\right)
\right\} &=& \left(\Phi_2(\sigma ) +\Phi_2(\sigma^\prime )\right)\partial_\sigma\delta\left(\sigma -\sigma^\prime \right)\nonumber\\
\left\{ \Phi_{1}\left(\sigma\right),\Phi_{1}\left(\sigma^{\prime}\right)
\right\} &=& \left(\Phi_1(\sigma ) +\Phi_1(\sigma^\prime )\right)\partial_\sigma\delta\left(\sigma -\sigma^\prime \right)\nonumber\\
\left\{ \Phi_{2}\left(\sigma\right),\Phi_{2}\left(\sigma^{\prime}\right)
\right\} &=& \left(\Phi_1(\sigma ) +\Phi_1(\sigma^\prime )\right)\partial_\sigma\delta\left(\sigma -\sigma^\prime \right)
\label{constalng1}  
\end{eqnarray}

Let us next perform a degree of freedom count. The total number of phase space degrees of freedom is $2(6+k)$. There are $4$ second class constraints and $4$ first class constraints. Hence the number of independent phase space degrees of freedom is,
\begin{equation}
n=2(6+k) - 4 - 2\times 4 = 2k
\label{dofcount}
\end{equation}
Hence the independent number of configuration space degrees of freedom is $k$, which we take to be the $X^k$ variables. This precisely matches wih our earlier counting and identification.

Since the constraints in (\ref{psc}) are strongly implemented, it is possible to solve for $\alpha$ and $\beta$. We get,
\begin{eqnarray}
\alpha = \Pi^0 X^{\prime 1 } + \Pi_1 X^{'0}  \nonumber\\
\beta  = -\Pi^0 X^{\prime 0 } -  \pi_1X^{\prime 1 }
\end{eqnarray} Substituting 
these in (\ref{interconstraints}) we obtain,
\begin{equation}
\Phi_1 = \Omega_1 \,\,\, ; \,\,\, \Phi_2 = \Omega_2
\end{equation}
 which are identical to the  two first class constraints (\ref{115}) of the original Nambu - Goto model.

This shows that the interpolating model is perfectly viable, allowing for a systematic analysis of constraints that is needed to show its equivalence with  other types of actions. This hamltonian analysis complements the lagrangian fomulation presented previously.

 \section{Connection with geometry}
 
   In the above we have introduced a new action for the nonrelativistic bosonic string
which has the merit of interpolating between the Nambu Goto form on one side and the Polyakov form of the action on the other. Thus both types of actions can be related in one go. In the nonrelativistic variety this task is not simple, as evidenced in the literature \cite{BGN, GOY}. First of all, metric components in the transverse directions only appear in the Polyakov type action and there is a clear compartmentalization of the directions horizontal to the string and transverse to  the string in the target space. The low energy excitations are entirely transverse. The horizontal components of the metric are not dynamical. Probably due to this proviso two new fields are included in the action which are devoid of any dynamics \cite{BGN}. How these fields are related with the geometry (i.e. their connection with the metric)  is not known. Clearly the action level correspondence between the different forms of the action, so transparent in the relativistic formulations, appear to be missing. The equivalence could be established
by an ardous path \cite{BGY}. The new action found here not only was shown to  bridge the different forms, it also generated the additional fields. 
 Moreover it has connected the geometric elements with the multipliers in the Hamiltonian.
 In the following we will further investigate the connection  and one would appreciate that this connection is not accidental.


   So far, the metric induced on the world sheet was not discussed  because the discussion could proceed  without  reference to the metric. A metric on the world sheet was derived as the metric induced by the embedding (see equation (\ref{ngrel})). But the Polyakov form gives an independent metric. So
geometry of the world sheet is now interwoven with the dynamics.
In deriving the Polyakov type action from the Nambu Goto through the interpolating action, we have connected the form of the metric (\ref{adm}) composed with the Hamiltonian constructs  $\rho$ and $\lambda $, with the world sheet geometry. This evolution of the metric as dynamical fields is certainly a new input in the existing literature.


 Our construction (\ref{adm}) is reminiscent of the Arnowitt--Deser--Misner (ADM) decomposition of general relativity \cite{MTW}.
In the ADM representation the metric of the four dimensional Riemannian space time ${}^{\left(4\right)}\gamma^{\mu\nu}$ is split as\footnote{For the metric of the total space time the dimension is mensioned as a (pre)superscript} 
\begin{equation}
\left(\begin{array}{cc}
{}^{\left(4\right)}\gamma^{00}&{}^{\left(4\right)}\gamma^{0m}\\
{}^{\left(4\right)}\gamma^{0k}&{}^{\left(4\right)}\gamma^{km}
\end{array}\right)
 =\left(\begin{array}{cc}
-\frac{1}{\left(N\right)^{2}}&\frac{\left(N^{m}\right)}{\left(N\right)^{2}}\\
\frac{\left(N^{k}\right)}{\left(N\right)^{2}}&\left(\gamma^{km}-\frac{\left(N^{k}\right)\left(N^{m}\right)}{\left(N\right)^{2}}\right)
\end{array}\right)
\label{ADM4}
\end{equation}
Here, $k$, $m$ take the values $1$, $2$, $3$. $\gamma^{km}$ is the metric on a three dimensional hypersurface embedded in the four dimensional space time. $N$ and $N^{k}$ are the arbitrary lapse and shift variables which are nothing but the Lagrange multipliers of the theory. From (\ref{ADM4}), a similar structure for $d = 2$ assumes the following form 
\begin{equation}
\left(\begin{array}{cc}
{}^{\left(2\right)}\gamma^{00}&{}^{\left(2\right)}\gamma^{01}\\
{}^{\left(2\right)}\gamma^{01}&{}^{\left(2\right)}\gamma^{11}
\end{array}\right)
 =\left(\begin{array}{cc}
-\frac{1}{\left(N\right)^{2}}&\frac{\left(N^{1}\right)}{\left(N\right)^{2}}\\
\frac{\left(N^{1}\right)}{\left(N\right)^{2}}&\left(\gamma^{11}-\frac{\left(N^{1}\right)^{2}}{\left(N\right)^{2}}\right)
\end{array}\right)
\label{ADM2}
\end{equation}
Comparing this with (\ref{adm}) we can easily establish 

\begin{eqnarray}
N^{1}\mapsto \rho \quad \mathrm {and}\quad \left(N\right)^{2}\mapsto -\lambda\sqrt{-g}\quad \mathrm{and}\quad \gamma\mapsto g
\label{ADM3}
\end{eqnarray}
Thus the  identification (\ref{adm}) is the same as the ADM foliation of the world-sheet. 
 The fields $\lambda$ and $\rho $ are  manifestations of the arbitrariness along the string (the lapse $N$) and an arbitrariness transverse to the string i.e. the relative time direction (the shift $N^1$), both on the world sheet.

  For furtner insight and comparison with existing results \cite{GOY}, we rewrite the interpolating Lagrangian in 
    the light-cone coordinates,
\begin{equation}
X = X^0 + X^1 \qquad \bar{X} = X^0 - X^1
\end{equation} 
The extra piece can now be written as,
\begin{eqnarray}
{\mathcal {L}_e} = \alpha \left( \frac{1}{\sqrt{-h}h^{00}} + \frac{\epsilon^{\alpha\beta}\partial_{\alpha }X \partial_{\beta}{ \bar{X}}}{{2{\bar{X}}^\prime}X^{\prime} }\right)  + \beta \left(-\frac{h^{01}}{h^{00}} - \frac{\sigma^{\alpha \beta}\partial_\alpha X\partial_\beta \bar X}{2X^{ \prime}\bar{X}^{ \prime} } \right)
\label{extra}
\end{eqnarray}
where ,
\begin{equation}
\epsilon^{\alpha\beta} =
\left(\begin{array}{cc}
0 & 1 \\
-1 & 0\\
\end{array}\right)
\label{epsilon}
\end{equation}
and $\sigma^{\alpha\beta}$ is
\begin{equation}
\sigma^{\alpha\beta} =
\left(\begin{array}{cc}
0 & 1 \\
1 & 0\\
\end{array}\right)
\label{sigma}
\end{equation}
Combining (\ref{main}) and (\ref{extra}) we get the interpolating action in light cone coordinates
\begin{equation}
{\mathcal{ L_I}} = \frac{1}{2}\sqrt{-h}h^{\alpha\beta}\partial_\alpha X^k \partial_\beta X^k + \alpha \left( \frac{1}{\sqrt{-h}h^{00}} + \frac{\epsilon^{\alpha\beta}\partial_\alpha X\partial_\beta {\bar{X}}}{{2{\bar{X}}^\prime}X^{\prime} }\right)  + \beta \left(-\frac{h^{01}}{h^{00}} - \frac{\sigma^{\alpha \beta}\partial_\alpha X\partial_\beta \bar X}{2X^{ \prime}\bar{X}^{ \prime} } \right) 
\label{test}
\end{equation}

 Now note that the coordinates $X^a$ are defined in a Lorentz plane with metric $\eta_{ab} = (1, -1)$. The string world sheet, in its ground state is parallel with the Lorentz plane. When the world sheet metric is an independent field, one has to introduce the  tangent space at every  point on the world sheet. The tangent space is  locally Lorentzian. The coordinates, $X^0$ and $X^1$ are referred to these coordinates. Let $e_\alpha$ and $e_a$ be the bases at a point on the world sheet and the tangent space at that point, respectively. The vierbein ${\Lambda^{\alpha}}_{a}$ and its inverse connect the two bases,
 \begin{equation}
 e_\alpha = \Lambda_\alpha{}^a e_a.
 \end{equation}
 The inverse of $\Lambda_\alpha{}^a$ will be denoted by $ \Lambda^\alpha{}_a $.
 
 The vierbeins may be used to factorize the metric,

\begin{equation}
h ^{\alpha \beta} = {\Lambda^{\alpha}}_a{\Lambda^{\beta}}_b\delta^{ab}
\label{vierbein}
\end{equation} 

We now give special attention to the part of the Lagrangian where the vierbeins explicitly
appear in the theory. It is that part of the Lagrangian which is special to the non relativistic theory having no relativistic analog. Now, both $X$
 and $\bar{X}$ are world sheet scalars. So 
we can easily derive the following relations
\begin{eqnarray}
\partial_{\alpha} X &=& {\Lambda_\alpha}^a\partial_a X = {\Lambda_\alpha}^a\left( \delta_a{}^0
+ \delta_a^1 \right) = e_\alpha\nonumber\\
\partial_{\alpha} {\bar{X}} &=& {\Lambda_\alpha}^a\partial_a {\bar{X}} = {\Lambda_\alpha}^a\left( \delta_a{}^0
- \delta_a^1 \right) = {\bar{e}}_\alpha
\label{relations}
\end{eqnarray} 
where 
 
\begin{eqnarray}
e_\alpha &=& \Lambda_{\alpha}^0 + \Lambda_{\alpha}^1\nonumber\\
{\bar{e}_\alpha} &=& \Lambda_{\alpha}^0 - \Lambda_{\alpha}^1
\label{e}
\end{eqnarray}

Now we have all the intermediate quantities. We can then write the expression in terms of the basis vectors,
\begin{equation}
{\mathcal{ L_I}} = \frac{1}{2}\sqrt{-h}h^{\alpha\beta}\partial_\alpha X^k \partial_\beta X^k + \alpha \left( \frac{1}{\sqrt{-h}h^{00}} + \frac{\epsilon^{\alpha\beta}e_\alpha \bar e_\beta}{{2{\bar{X}}^\prime}X^{\prime} }\right)  + \beta \left(-\frac{h^{01}}{h^{00}} - \frac{\sigma^{\alpha \beta}e_\alpha \bar e_\beta} {2X^{ \prime}\bar{X}^{ \prime} } \right) 
\label{test1}
\end{equation} 

The string action in the form (\ref{test1}) clearly reveals the connection with the world sheet geometry. The deduction of this from the canonical analysis of the model (\ref{1000}) is indeed remarkable.

 



\section{Conclusion}
Nonrelativistic string theories (NRST) have recently come to prominence in the literature \cite{GO, BGN, BGY, GOY}. Just as recent studies of nonrelativistic field theories have emphasised the role of geometry to tackle the issue of coupling  nonrelatividtic field theories with gravity, studies in  string theories have raised new questions about geometry. But nay, the NRSTs are interesting even in flat space.
Similar to their relativistic counterpart, different actions have been proposed. Broadly we can divide these in two classes -- minimal action which contains the world sheet area swept by the string as the Lagrangian \cite{BGN} and more redundant form of action where
the metric elements on the world sheet \cite{GOY} are considered as independent fields. The former is comparable with the Nambu - Goto
type and the latter with the Polyakov type in relativistic strings. But there is one significant difference. In NRSTs in the second type there are two extra fields on the world sheet. In relativistic field theories of Polyakov type one can easily substitute
the metric components from their equations of motion to get the Nambu - Goto string. This action level correspondence is not apparent in case of the NRST primarily due to the presence of the extra fields. 

 We have developed
an approach where the starting point was the Nambu-Goto type of action. This was obtained by taking a nonrelativistic $(c\rightarrow \infty)$ limit of the relativistic Nambu-Goto action. A unique form for the action was found. By eliminating its gauge freedom it was shown to reduce to the expected form for the nonrelativistic stretched string vibrating tranversally. The choice of gauge was justified by a hamiltonian analysis. 

This hamiltonian analysis was pushed in a big way by constructing from the Nambu-Goto type action an intermediate action that interpolated between the Nambu-Goto and Polyakov type forms. While the connection with the Nambu-Goto form was easily shown, that with the Polyakov type was much more involved. It was possible to identify a metric structure using the fields introduced in the hamiltonian fomulation.  This enabled us to connect the string world sheet geometry with the geometry of the embedding which can be identified with that appearing in the Polyakov type action.
Thereby, we regain the action level correspondence in NRSTs. This shows the utility of our hamiltonian approach and explains the problems of showing    the equivalence by strictly confining to the lagrangian approach, as done for the relativistic example.

  Another remarkable aspect is the role of Hamiltonian analysis in the formulation of the
interpolating  action. We have started from the minimal action. To facilitate the introduction of geometry on the world sheet of the string we have carried out a comprehensive canonical analysis of the model. The action is then enriched by the introduction of the constraints in the Lagrangian by the Lagrange multiplier technique and lifting the status
of the multipliers to independent fields. Eliminating the  phase space variables by the inverse Legendre procedure, the desired action is obtained. A surprising connection of the new fields with the Arnowit - Deser - Misner construction in general relativity emerged, whereby geometry was introduced in the theory. It was then an easy journey towards the Polyakov type action. Remarkably, the two extra fields appeared
spontaneously in the process. 
We have provided a detailed canonical analysis of the new action. Its phase space structure has been studied. Throughout the paper symmetries of the different actions have been investigated  from the canonical point of view and the interpolating Lagrangian is no exception. This analysis has been used to deduce further geometrical connections.

 The interplay of canonical analysis and geometry, as evidenced here,  brings out a clear picture of the connection of canonical analysis with the metric of the world sheet. Thus  the possible use of the action obtained here in case of  coupling with gravity appears to be feasible. One has to find out  ways to relate the background curvature.  Our approach is amenable to generalisation and can be extended to membranes and higher branes. This is because the analysis has a close parallel with the relativistic case where the equivalence of higher brane actions was demonstrated  \cite{BMS1} by generalising the results for the string case done in \cite{BMS}.  This could be useful since the equivalence of actions for nonrelativistic theories is obviously much more complicated than its relativistic counterpart. These and other issues may yield fresh insights  and open up  further research.

\end{document}